\documentclass{ws-procs961x669}            

\usepackage{graphicx}	
\usepackage{multirow,makecell}

\RequirePackage{bm,amsfonts,hhline,amsmath,amssymb,microtype,float,eurosym,
latexsym,epsf,mathtools,cuted,times,makecell,array,natbib,cancel}
\usepackage{tabularx}
\usepackage{epstopdf}

\usepackage{gensymb}
\usepackage{xcolor}

\usepackage{pifont}

\usepackage{hyperref}

\begin{document}
\title{Gravitational geometric phase}

\author{Banibrata Mukhopadhyay}
\address{Department of Physics, Indian Institute of Science, Bangalore 560012, India\\
E-mail: bm@iisc.ac.in}
\author{Tanuman Ghosh}
\address{Raman Research Institute, Bangalore, 560080, India\\
E-mail: tanuman@rri.res.in}
\author{Soumya Kanti Ganguly}
\address{Ongil Private Limited, Tidel Park, Rajiv Gandhi IT Expressway,
Chennai 600113, India\\
E-mail: soumya09ganguly@gmail.com}

\begin{abstract}
We show that spinors propagating in curved gravitational background
acquire an interaction with spacetime curvature, which leads to a 
quantum mechanical geometric effect. This is similar to what happens
in the case of magnetic fields, known as Pancharatnam-Berry phase. 
As the magnetic and gravitational fields have certain similar
properties, e.g. both contribute to curvature, this result is not
difficult to understand. Interestingly, while spacetime around a 
rotating black hole offers Aharonov-Bohm and Pancharatnam-Berry both kinds of geometric effect, a static spacetime offers only the latter.
In the bath of primordial black holes, such gravity induced effects 
could easily be measured due to their smaller radius.
\end{abstract}


\bodymatter

\section{Introduction}\label{SecI}
In recent works \cite{Mukhopadhyay1,TM}, we showed that spinors propagating in a curved background acquire a geometric phase (GP) similar to the case of GP in electromagnetic field.  The study of GP in electromagnetic field has been widely explored. Two most important notions of GPs are: Aharonov-Bohm (AB) effect \cite{AB1} and Pancharatnam-Berry (PB) \cite{pancharatnam1, Berry1} phase. AB effect and PB phase occur in two different physical scenarios. In presence of a varying magnetic field, PB phase is acquired by a spinor, whereas AB effect can be acquired even without a varying field. In fact, AB effect is one of the signature properties of magnetic potential influencing a particle's quantum state. These effects/phases in case of electromagnetic fields originate from the effective curvature of magnetic potential and field in the (flat) Minkowski spacetime. We showed \cite{TM} that in a curved spacetime, a spinor in the vicinity of a gravitating body can acquire similar geometric phases which originate from the effective curvature of the spacetime itself.

Analogous effects for gravitational and magnetic fields are well known. Some of the examples are: curvature generated in both scenarios; electromagnetic radiation from accelerated charge and gravitational radiation from accelerated mass (only quadrupolar in nature), both propagating with the speed of light; energy splitting of spin-half particles in presence of both fields \cite{Parker1,book}. With the continuation of these analogous results, it is expected that the quantum effects, which generate the GPs in case of magnetic field, similarly should be present in case of gravitational field. Some studies have already explored these effects in local coordinates \cite{Mukhopadhyay1,book,Dixit1}. There are many astrophysical and cosmological phenomena like baryogenesis, neutrino emissions from supernovae and active galactic nuclei (AGNs), neutrino oscillation, neutrino dominated accretion disks, Fermi degenerate gas in compact objects, etc., where study of spinor propagation and evolution is crucial to understand the physical nature of these phenomena \cite{Mukhopadhyay2,Mukhopadhyay3,
Kosteleck1,Piriz1,Chen1,Cardall1,Das1,Cook1}.

When it comes to the study of gravity with quantum mechanics and understand various effects of quantum systems in curved spacetime, semi-classical formalism is a very effective tool to perform such study. In fact, there are plenty of available works in the literatures which have explored the dynamics of spinors in curved spacetime. These have been done with the Lagrangian \cite{sch,Mukhopadhyay2} and Hamiltonian \cite{Parker1,Huang1} formalisms. It is also well known fact that in the Hamiltonian formalism, the Dirac Hamiltonian of a spinor in curved spacetime has a non-hermiticity and uniqueness problem. Different authors tried to implement different approaches to address these issues \cite{Parker1,Huang1,Obukhov1,Obukhov2,Gorbatenko1, Gorbatenko2}. Non-hermitian biorthogonal quantum mechanics \cite{Brody_2013} provides tool for solving the non-hermiticity issue of Dirac Hamiltonian. One approach is to define a relativistically invariant scalar product (Parker scalar product) \cite{Parker1, Huang1} and use that scalar product to define the expectation values of all operators. In case of curved spacetime, this scalar product is different from the standard flat spacetime scalar product because of the presence of curved spacetime metric and vierbeins. The other approach is known as $\eta-$ representation or pseudo-Hermitian approach \cite{Gorbatenko1, Gorbatenko2}, where the Hamiltonian and eigenstates of the particle are modified by the metric and vierbeins in such a way that effectively one can use the standard flat scalar product to define all the operators in the Hilbert space. In fact, these two approaches are shown to be equivalent \cite{Gorbatenko1, Gorbatenko2}.

In this work, we explore the Dirac Hamiltonian in curved spacetime background, especially in Kerr and Schwarzschild geometries, using the pseudo-Hermitian approach or the $\eta$-representation and explore the possibilities of appearance of GPs in spinors traversing in curved spacetime. In section \ref{SecII}, we recapitulate the quantum mechanics in $\eta$-representation. Then in section \ref{SecIII}, we derive the Dirac Hamiltonian in the Kerr geometry and further derive 
its non-relativistic counterpart for slowly moving particles around
a weakly rotating black hole and its analogy
with that in electromagnetic fields in section \ref{SecIV}. In section \ref{SecV}, we find GPs in curved spacetime. Finally, in section \ref{SecVI}, we summarize our findings.

\section{Quantum Mechanics in the $\eta$-Representation}\label{SecII}
We briefly discuss the quantum mechanics in $\eta$-representation \cite{Gorbatenko1,Gorbatenko2,Bender1,Mostafazadeh1}. If an operator $\eta$ satisfies the relation 
\begin{equation}
\rho = \eta^{\dagger} \eta,
\label{eta1}
\end{equation}
where $\rho$ is an invertible operator which assures the relation to be satisfied as
\begin{equation}
\rho H \rho^{-1} = H^{\dagger},
\end{equation}
then the Hamiltonian in $\eta$-representation turns out to be
\begin{equation}
H_{\eta} = \eta H \eta^{-1} = H_{\eta}^{\dagger}.
\end{equation} 

The relation between the wave functions in the $\eta$--representation and in the standard representation is given by
\begin{equation}
\Psi = \eta \psi,
\label{eta2}
\end{equation}
where these two wave functions satisfy the following wave equations:
\begin{equation}
i \frac{\partial \psi}{\partial t} = H \psi,
\label{Schro_eq1}
\end{equation}

\begin{equation}
i \frac{\partial \Psi}{\partial t} = H_{\eta} \Psi.
\label{Schro_eq2}
\end{equation}

As we mentioned above, the Parker scalar product \cite{Parker1,Huang1}, which is a modification of standard scalar product, in Hilbert space is defined as
\begin{equation}
(\phi, \psi)_{\rho} = \int d^3x (\phi^{\dagger} \rho \psi),
\label{scalar_product1}
\end{equation}
whereas in the pseudo-Hermitian approach, the scalar product takes the  standard form of flat space scalar product since the Hamiltonian and the wave functions are already modified there, which is

\begin{equation}
(\Phi, \Psi) = \int(\Phi^{\dagger} \Psi)d^3x.
\label{scalar_product2}
\end{equation}
We can easily verify that
\begin{equation}
(\phi, \psi)_{\rho} = (\Phi, \Psi).
\end{equation}

\section{Dirac Hamiltonian in the Kerr and Schwarzschild metrices}\label{SecIII}
We use the convention of natural unit system with $\hbar=G=c=1$ and metric 
signature of $(+,-,-,-)$.

We can start from the Dirac equation in curved spacetime which is,
\begin{equation}
(i \gamma^{\mu} D_{\mu} - m) \psi (x) = 0.
\label{Dirac equation1}
\end{equation}
where the covariant derivative $D_{\mu} = \partial_{\mu}+\Gamma_{\mu}$. Here $\Gamma_{\mu}$ is the spinorial affine connection \cite{Parker1,Huang1}. $m$ is the mass of the Dirac particle, $\psi$ is the four-component column bispinor. $\gamma^{\mu}$ are the Dirac matrices in curved spacetime which satisfies the relation
\begin{equation}
\gamma^{\alpha}\gamma^{\beta}+\gamma^{\beta}\gamma^{\alpha} =  2 g^{\alpha \beta} I_4,
\end{equation}
where $I_4$ is the $4 \times 4$ identity matrix and $g^{\alpha\beta}$ is the contravariant metric tensor of curved spacetime. The adjoint spinor is defined as $\bar{\psi} = \psi^\dagger \gamma^0 $.

The gamma matrices in curved spacetime, also known as global gamma matrices ($\gamma^{\alpha}$), are related to the local flat spacetime gamma matrices ($\gamma^{a}$) by the relation 
\begin{equation}
\gamma^{\alpha} = e^{\alpha}_{a} \gamma^{a},
\label{tetrad_relation}
\end{equation}
where $e^{\alpha}_{a}$-s are the tetrads defined by 
\begin{equation}
g_{\mu \nu} = e^{a}_{\mu} e^{b}_{\nu} \eta_{ab}.
\end{equation}
We then derive the global Dirac Hamiltonian operator from equation (\ref{Dirac equation1}), which is
\begin{eqnarray}
H = - i \Gamma_{t}- i (g^{tt})^{-1}\gamma^{t}  \left[\gamma^r (\partial_r +\Gamma_r)\right. \nonumber \\
\left. +\gamma^{\theta} (\partial_{\theta} + \Gamma_{\theta}) +\gamma^{\phi} (\partial_{\phi} + \Gamma_{\phi})\right] + (g^{tt})^{-1} \gamma^t m. 
\end{eqnarray}
This Hamiltonian is self-adjoint under the Parker scalar product given in equation (\ref{scalar_product1}). Nevertheless, our purpose is to find the Dirac Hamiltonian in $\eta$-representation which will be useful for working in standard flat Hilbert space. We choose astrophysically important metrices to work with and hence we shall find the Dirac Hamiltonian in $\eta$-representation for the Kerr metric, which reduces to 
the Schwarzschild metric when the Kerr parameter $a=0$. 

\subsection{Kerr metric}
\label{Kerr}

The Kerr metric in the Boyer-Lindquist coordinates is
\begin{align}
ds^2 = \left(1-\frac{2Mr}{\rho^2}\right)dt^2 + \frac{4Mra \sin^2 \theta}{\rho^2} dt d\phi - \frac{\rho^2}{\Delta} dr^2 \nonumber \\
 - \rho^2 d\theta^2 - \left[(r^2+a^2)\sin^2\theta + \frac{2Mra^2 \sin^4\theta}{\rho^2}\right] d\phi^2, 
\label{kerr_metric}
\end{align}
where $\rho^2 = r^2+a^2 \cos^2\theta$ , $\Delta = r^2-2Mr+a^2$, $M$ is the mass
of black hole and $a$ is the Kerr parameter (angular momentum per unit mass of
the black hole). We choose the Schwinger gauge of tetrad \cite{Schwinger1,Gorbatenko3,Neznamov1}, given by

\begin{eqnarray}
e^t_0=\sqrt{g^{tt}} , e^r_1 = \frac{\sqrt{\Delta}}{\rho} , e^\theta_2 = \frac{1}{\rho} , \nonumber \\ e^\phi_3 = \frac{1}{\sin\theta \sqrt{\Delta}\sqrt{g^{tt}}} , e^\phi_0 = \frac{2Mar}{\rho^2\Delta\sqrt{g^{tt}}}.
\end{eqnarray}
The Dirac Hamiltonian in the $\eta$-formalism in Kerr metric then turns out to be \cite{Gorbatenko4}

\begin{align}
H_\eta = \frac{m}{\sqrt{g^{tt}}} \gamma^0 - i \frac{\sqrt{\Delta}}{\rho \sqrt{g^{tt}}} (\frac{\partial}{\partial r} + \frac{1}{r}) \gamma^0 \gamma^1 \nonumber \\
- i \frac{1}{\rho \sqrt{g^{tt}}} (\frac{\partial}{\partial \theta}+\frac{1}{2} \cot\theta) \gamma^0\gamma^2 -i \frac{1}{g^{tt}\sqrt{\Delta}\sin\theta}\frac{\partial}{\partial \phi} \gamma^0\gamma^3 \nonumber \\
- i \frac{2Mar}{g^{tt}\rho^2\Delta}\frac{\partial}{\partial \phi} - i \frac{1}{2}\frac{\partial}{\partial r}(\frac{\sqrt\Delta}{\rho\sqrt{g^{tt}}}) \gamma^0 \gamma^1 -i \frac{1}{2}\frac{\partial}{\partial \theta}(\frac{1}{\rho\sqrt{g^{tt}}}) \gamma^0 \gamma^2 \nonumber \\
+ i \frac{{\sqrt{g^{tt}} \Delta M a \sin \theta}}{2 \rho} (\frac{\partial}{\partial r} (\frac{r}{g^{tt}\rho^2\Delta}) \gamma^3 \gamma^1 \nonumber \\
+\frac{1}{\sqrt{\Delta}} \frac{\partial}{\partial \theta} (\frac{r}{g^{tt}\rho^2\Delta}) \gamma^3 \gamma^2).
\label{Hk}
\end{align}

We can write this Hamiltonian as
\begin{eqnarray}
H_\eta = ({\sqrt{g^{tt}}})^{-1}\left[\gamma^0 m + \gamma^0 \gamma^j (p_j-i A_j)
	+ i\gamma^0\gamma^j\gamma^5 k_j + e^{\phi}_{0} p_{\phi}\right],
\label{hamil_kerr}
\end{eqnarray}
where $A_1= \frac{\sqrt{\Delta}}{\rho r}+\frac{\sqrt{g^{tt}}}{2}\frac{\partial}{\partial r}\left(\frac{\sqrt{\Delta}}{\rho \sqrt{g^{tt}}}\right)$, $A_2 = \frac{\cot \theta}{2 \rho}+\frac{\sqrt{g^{tt}}}{2}\frac{\partial}{\partial \theta}\left(\frac{1}{\rho \sqrt{g^{tt}}}\right)$ and $A_3 = 0$; 
$k_1 =  \frac{ig^{tt}M a \sqrt{\Delta} \sin\theta}{2 \rho} \frac{\partial}{\partial \theta} \left(\frac{r}{g^{tt}\rho^2\Delta}\right)$, $k_2 =  -\frac{i g^{tt}M a \Delta \sin\theta}{2 \rho} \frac{\partial}{\partial r}\left(\frac{r}{g^{tt}\rho^2 \Delta}\right)$ and $k_3=0$.
Here we use the fact that $\gamma^3\gamma^2$ and $\gamma^3\gamma^1$ can be respectively written as $i\gamma^0\gamma^1\gamma^5$ and $-i\gamma^0\gamma^2\gamma^5$.

Now using the relations $H_\eta = i \frac{\partial}{\partial t}= p_t$ and $ \partial_0 = e^t_0 \partial_t + e^{\phi}_0 \partial_\phi$, we rearrange the equation (\ref{hamil_kerr}) to
\begin{eqnarray}
\left[p_0 - \gamma^0\{m + \gamma^j (p_j-i A_j) + i\gamma^j \gamma^5 k_j\}\right] \Psi = 0,
\label{hamil_kerr1}
\end{eqnarray}
where we use the tetrad transformation relation between global and local 4-momenta as 
\begin{equation}
p_j = e^{\mu}_j p_{\mu},
\end{equation}
where Greek indices represent global coordinates ($t,r,\theta,\phi$) and  roman indices represent flat coordinates ($0,1,2,3$).

Now we need to identify the terms in these equations with the analogy of magnetic field scenario which we discussed in section \ref{SecI}. $A_j$s are analogous to the magnetic vector potential and $k_j$s are ``pseudo-vector" potential, which in this case of Kerr metric appears due to the chirality in the system owing to the rotation of spacetime.

If we put $a=0$, the equations in the Kerr metric reduce to those in the Schwarzschild metric.
The Hamiltonian then turns out to be 
\begin{align}
	H_\eta = ({\sqrt{g^{tt}}})^{-1}\left[\gamma^0 m + \gamma^0 \gamma^j (p_j-i A^s_j)\right],
\label{hamil_schwarz}
 \end{align}
where $A^s_1 = \frac{1}{r}\sqrt{1-\frac{2M}{r}} + \frac{\sqrt{g^{tt}}}{2}\frac{\partial}{\partial r} (\frac{\sqrt{1-\frac{2M}{r}}}{\sqrt{g^{tt}}})$ , $A^s_2 = \frac{\cot\theta}{2r}$ and $A^s_3 = 0$.

We can also rearrange the Hamiltonian for $a=0$ to write the Dirac equation in a compact form as
\begin{align}
\left[p_0 -\gamma^0\{m + \gamma^j (p_j-i A^s_j)\}\right] \Psi = 0.
\label{hamil_scharz1}
\end{align}
Here $A^s_j$s are the ``gravito-magnetic potential" in the Schwarzschild geometry similar to the $A_j$s in Kerr metric.

\section{Nonrelativistic Approximation of Dirac Hamiltonian}\label{SecIV}
In the standard Dirac representation, we write the Dirac equation in Kerr metric from equation (\ref{hamil_kerr1}) as
\begin{equation}
  \begin{pmatrix}
  -p_0+i  \vec{\sigma} \cdot \vec{k}  +m & \vec{\sigma}\cdot\vec{\Pi}_A \\
     \vec{\sigma}\cdot\vec{\Pi}_A &-p_0+i  \vec{\sigma} \cdot \vec{k}  -m
  \end{pmatrix} \Psi = 0,
 \end{equation}
 where
\begin{eqnarray}
\vec{k}=(k_1,k_2,0), \nonumber \\
\vec{\Pi}_A=\vec{p}-i\vec{A}.
\end{eqnarray}

Following the standard process of deriving the non-relativistic limit of the Hamiltonian (see reference \refcite{sakurai1967advanced}) we can write the coupled equations as
\begin{subequations}
\begin{align}
(\vec{\sigma}\cdot \vec{\Pi}_A) \Psi_B = (E-i\vec{\sigma} \cdot \vec{k} - m) \Psi_A,  \label{eq:subeq1} \\
(\vec{\sigma}\cdot \vec{\Pi}_A) \Psi_A = (E-i\vec{\sigma} \cdot \vec{k} + m) \Psi_B. \label{eq:subeq2}
\end{align}
\end{subequations}
where $E$ is the energy eigenvalue, $\Psi_A$ and $\Psi_B$ are the two components of wavefunction $\Psi$ and we take into account the fact that $\vec{A}$ and $\vec{k}$ are time-independent. 

We now assume slowly rotating spacetime and non-relativistic particles 
such that the particle velocity $v<<c$, hence
\begin{equation}
E \sim m, \hspace{0.5cm} |\vec{k}|<<m 
\end{equation}
and define the non-relativistic particle energy,
\begin{equation}
E^{NR} = E - m.
\end{equation}

It can easily be derived that by keeping only leading order terms by
combining equations (\ref{eq:subeq1}) and (\ref{eq:subeq2}), we obtain
\begin{eqnarray}
\frac{1}{2m} (\vec{\sigma}\cdot\vec{\Pi}_A)(\vec{\sigma}\cdot\vec{\Pi}_A) \Psi_A = (E^{NR} - i\vec{\sigma}\cdot\vec{k}) \Psi_A,
\end{eqnarray}
which can be further written as
\begin{eqnarray}
\left[\frac{\Pi^2}{2m} + \sigma \cdot (i\vec{k}-\frac{i}{2m}\vec{B}_g)\right] \Psi_A = E^{NR} \Psi_A,
\label{Hamil_NR_kerr1}
\end{eqnarray}
where $\vec{B}_g=\vec{\nabla}\times\vec{A}$ is the effective ``gravito-magnetic field" and $\vec{A}$ is ``gravito-magnetic potential". This implies that
\begin{equation}
H^{NR} \Psi_A =\left[\frac{\Pi^2}{2m} + \vec{\sigma} \cdot \vec{B}^{kerr}_g\right] \Psi_A, 
\label{Hamil_NR_Kerr2}
\end{equation}
where $\vec{B}^{kerr}_g = (i\vec{k}-\frac{i}{2m}\vec{B}_g)$ which involves both field analogue term and potential analogue term.

Similarly for $a=0$ (Schwarzschild) case, we can find that
\begin{equation}
{H^s}^{NR} \Psi_A  =\left[\frac{{\Pi^s}^2}{2m} + \vec{\sigma} \cdot \vec{B}^{sch}_g\right] \Psi_A,
\label{Hamil_NR_Sch1}
\end{equation}
where $\vec{B}^{sch}_g = -\frac{i}{2m}\vec{B}^{s}_g$ where $\vec{B}^{s}_g = \vec{\nabla}\times\vec{A^s}$.

Note importantly that $\vec{B}^{kerr}_g$ involves the field ($\vec{B}_g$) 
and potential ($\vec{k}$) both, whereas $\vec{B}^{sch}_g$ involves only field.

\subsection{Analogy with electromagnetism}

In the presence of electromagnetic field, the Dirac equation is given by
\begin{eqnarray}
\left[i\gamma^\mu\left(\partial_\mu-ieA_\mu\right)-m\right]\psi=0,
\label{direm}
\end{eqnarray}
where $e$ is the electric charge and $A_\mu$ is the electromagnetic covariant 
4-vector potential.
For the non-trivial solution for $\psi$, the~energies/Hamiltonians of the 
spin-up and spin-down particles are
given by
\begin{eqnarray}
(H+eA_0)^2=({\hat p}-e{\vec A})^2+m^2+e{\vec \sigma}\cdot{\vec B},
\label{direme}
\end{eqnarray}
where $A_0$ is the temporal component of $A_\mu$ which is basically the
Coulomb potential and $\hat{p}$ is the quantum mechanical momentum operator 
$-i\nabla$.
In the non-relativistic limit, equation (\ref{direme}) reduces to \cite{Mukhopadhyay1}
\begin{eqnarray}
H=-eA_0\pm\left[\frac{({\hat p}-e{\vec A})^2}{2m}+m+
\frac{e{\vec \sigma}\cdot{\vec B}}{2m}\right].
\label{direme2}
\end{eqnarray}
Clearly equation (\ref{direme2}) is very similar to equation 
(\ref{Hamil_NR_Sch1}) and, hence, whatever effects have been proposed for 
electromagnetism, the same are expected to suffice for gravitation. 

Apart from the split due to the positive and negative energy solutions, clearly
there is an additional split in the respective energy levels in equations
(\ref{Hamil_NR_Sch1}) and (\ref{direme2}). This is
basically the Zeeman-splitting (or Zeeman-like for gravitation) governed by the term
with Pauli's spin matrix,
in the up- and down-spinors for the positive and negative energy spinors
induced
by magnetic/gravitational fields, whether we choose relativistic or non-relativistic regimes.

\section{Geometric Phase}\label{SecV}
To find the geometric effects, we construct Poincar\'e sphere by $\vec{B}^{kerr}_g$.
Essentially, the interaction between the spinor and the gravitational background comes from the interaction Hamiltonian of
\begin{equation}
H_{int}=  \vec{\sigma} \cdot \vec{B},
\end{equation}
where $\vec{B}=\vec{B}^{kerr}_g$. We can write this Hamiltonian as

\begin{equation}
H_{int}= |\vec{B}|
  \begin{pmatrix}
     \cos\zeta &\sin\zeta \exp(-i\xi) \\
     \sin\zeta \exp(+i\xi) & -\cos\zeta 
  \end{pmatrix},
\end{equation}
where $\zeta$ and $\xi$ are the latitude and azimuthal angles of spherical polar coordinates respectively of the parameter space constructed by the vector $\vec{B}$. Here $\rho$ is the radial coordinate in this system.

Now, using the standard definition of phase ($\Phi_B$) and connection ($\vec{A}_B$) (see reference \refcite{TM} for details)
we find that the connection is
\begin{equation}
\vec{A}_B = \frac{(1-\cos\zeta)}{2\rho\sin\zeta} \hat{\xi},
\end{equation}
and the phase is
\begin{equation}
\Phi_B =\frac{ \tilde{\xi}}{2} (1-\cos\zeta)=\frac{\Omega}{2},
\label{equation_berry_phase}
\end{equation}
where $\tilde{\xi}$ is the total integrated azimuthal coordinate and
$\Omega$ is the integrated solid angle. 

Thus the GP in curved spacetime around Kerr (and Schwarzschild) geometry takes the similar form as revealed in the case of magnetic field but the angle coordinates and phases are fixed by the spacetime geometry.
It is important to note that in the Kerr metric case, the GP is a combination of AB effect which appears due to the potential like term $\vec{k}$ and PB phase which appears due to the field like term $\vec{\nabla} \times \vec{A}$. In case of Schwarzschild metric however only PB contributes to the GP.

\subsection{Possible measurement}
Previous studies \cite{Youlin1,Colella1} have proven the relevance of semi-classical effects in gravitational background for massive particles. In this paper, we have shown that origin of GPs in spinors traversing in  curved spacetime is theoretically possible. However, the question remains that how effectively these effects can be measured. For that, we need to estimate some of the length scales of the systems. Usually, quantum effects become prominent when length scale ($l$) of the system becomes comparable to or less than the de-Broglie wavelength of the particle $\lambda = \hbar/p$, where $p$ is the momentum. In the case of gravity, this $l$ is typically the radius of the gravitational body. We find that although the appearance of GPs is theoretically possible in all conditions, it 
may be detected only in selective cases like, e.g. primordial black holes (see reference \refcite{TM} for details). The qualitative reason is simple. Larger the size of the gravitating body, smaller the field $\vec{B}$ (or $\vec{A}$ or $\vec{k} \sim 1/r$) is and thus smaller the geometric effects are. Hence, for smaller black holes these effects are prominent, but not for astrophysical black holes

\section{Summary} \label{SecVI}
It is found that the Dirac particles interacting with spacetime curvature give rise to effective spin-orbit coupling which eventually manifests geometric 
phases/effects: Aharonov-Bohm (AB) effect and Pancharatnam-Berry (PB) phase. AB term originates from the spin of the background geometry which is responsible for the chirality in the system and this effect goes off when black hole spin becomes zero. PB phase appears in the Kerr metric as well as in spherically symmetric static spacetime like Schwarzschild geometry. Although theoretically these 
effects can appear in all cases, detection of such semi-classical effects is observationally possible, at least for nonrelativistic particles, in a scenario where a nonrelativistic massive particle moves around a primordial black hole. Future mission by {\it Fermi} satellite can prove the existence of primordial black holes by detecting small interference pattern within gamma-ray bursts. 
Although the results of this work are outcome for a special case of nonrelativistic particles and weakly rotating black holes, it is expected that a covariant formalism will provide similar results in a general gravitational background from the analogy of covariant formalism of study of GPs in magnetic fields \cite{Stone1}.


\end{document}